\documentstyle[12 pt]{article}
 
                \newcommand{\bea}{\begin{equation}}
                \newcommand{\eea}{\end{equation}}
                \newcommand{\ber}{\begin{eqnarray}}
                \newcommand{\eer}{\end{eqnarray}}
                \newcommand{\p}{\partial}
\begin{document}

                \title {\bf The Hydraulic Jump In Two Dimensions}
                \author {{\bf  Abhishek Mukherjee${}^1$, Amaresh Datta${}^1$}\\
                              {\bf   Jayanta K Bhattacharjee${}^2$}\\
                        ${}^1$ Department of Physics, Indian Institute of Technology,Bombay-400476\\
                        ${}^2$Department of Theoretical Physics, \\
                        Indian Association for the Cultivation of Science, \\
                        Jadavpur, Calcutta - 700032, India.\\
                        e - mail: tpjkb@iacs.ernet.in}
                \date{}
                \maketitle
                \begin{abstract}
                The first quantitative calculation of the position of the hydraulic 
                jump was carried out a few years ago by Bohr et.al. Since this is the only 
                calculation of the jump, we have analysed it from a slightly different 
                point of view. Our results are similar to those of Bohr et.al.
                \end{abstract}
                \newpage
                If a vertical jet of fluid impinges on a horizontal surface and spreads 
                out radially, then at a certain radius one observes a sudden jump in 
                the height of the fluid[1-4].It is a familiar observation,seen in the 
                kitchen sink everyday, but it has been a difficult system to deal with.The 
                question that is normal to ask is that given the volumetric flow rate 
                (or the Reynolds number of the impinging flow) can we predict the 
                position and magnitude of the jump which is known as the hydraulic jump.The 
                first attempt to provide a concrete answer to this question was 
                undertaken by Bohr,Dimon and Putkaradge[5] and took place only a few years 
                ago.
                \par 
                Continuity of mass flux and momentum flux is possible at any radius and 
                thus one more condition is needed to determine the radius of the 
                jump.This could be the energy flux but that is generally difficult to 
                handle.It was shown by Bohr et.al. that introducing viscosity does help in 
                setting up a criterion for finding the position of the jump. Since that 
                was the first attempt to set up a quantitative calculation, we believe 
                that it is worthwhile to look at the problem from a simpler standpoint.
                \par
                The central simplification in the calculation, a simplification also 
                used by Bohr et.al., is the assumption that the radial velocity 'u'is 
                greater than the vertical velocity 'w',while the variations with respect 
                to the vertical direction (ie.'z') is far more important than the radial 
                variations .With this in mind, the equation of continuity becomes:
                \ber
                \frac{\p}{\p r}(ur)+\frac{\p}{\p z}(wr)&=&0 \nonumber \\
                \mbox{or} \hspace{1.5cm}\frac{\p u}{\p r}+\frac{u}{r}+\frac{\p w}{\p 
                z}&=&0
                \eer
                The momentum flow equuation(Navier Stokes equation), for the radial 
                component,is:
                \bea
                u\frac{\p u}{\p r}+w\frac{\p u}{\p z}=-\frac{1}{\rho}\frac{\p p}{\p 
                r}+\nu\frac{\p ^2u}{\p z^2}
                \eea
                If Q is the volumetric flow rate,
                \bea
                Q=2\pi r\int_{0}^{h(r)}u(r,z)dz  ,
                \eea 
                where h(r) is the height of the fluid layer at distance r.The boundary 
                condition at z=0 is that the  velocity variations is constant with the 
                plate ie.u=0=w.On the surface {ie.z=h(r)}the stress is zero ie. 
                $\frac{\p u}{\p z}=0$.We now make the reasonable Ausatz that
                \bea
                u(r,z)= U(r) f(\eta)
                \eea
                where $\eta$ is the scaled variable $\eta=\frac{z}{h(r)}$.
                The boundary conditions decree that f(0)=0 ;f(1)=1 and 
                $f^(\prime)(1)=0$.From Eqn(3) we have  
                \ber
                Q &=& 2\pi r U(r) h(r) \int_{0}^{1} f(\eta) d\eta \nonumber \\
                &=& 2\pi r U(r) h(r) C
                \eer
                where $C= \int_{0}^{1} f(\eta) d\eta$.From Eqn(1) we have 
                \bea
                u(r,z)= \frac{Q}{2\pi C}\frac{f(\eta)}{r h(r)}
                \eea
                and
                \bea
                w(r,z)= U(r) h^{\prime}(r)\eta f(\eta)
                \eea
                Turning to Eqn(2), we see
                \ber
                u\frac{\p u}{\p r}&=&U(r) f \frac{\p}{\p r}(Uf)\nonumber \\
                &=&U U^{\prime} f^2+ U^2 f \frac{\p f}{\p \eta}\frac{\p \eta}{\p 
                r}\nonumber \\
                &=& U U^{\prime} f^2 - U^2 f \frac{\p f}{\p 
                \eta}\frac{h^{\prime}}{h}\eta
                \eer
                \ber
                w\frac{\p u}{\p z}&=& U h^{\prime}\eta f \frac{\p}{\p z}(Uf)\nonumber 
                \\
                &=& U^2 h^{\prime}\eta \frac{f}{h} \frac{\p f}{\p \eta}
                \eer
                \ber
                \nu\frac{\p^2 u}{\p z^2}&=& \nu\frac{\p^2 }{\p z^2}(Uf)\nonumber \\
                &=&\nu U \frac{\p }{\p z}(\frac{1}{h}\frac{\p f}{\p \eta})\nonumber \\
                &=&\nu U \frac{1}{h^2}\frac{\p^2 f}{\p \eta^2}
                \eer
                and
                \bea
                \frac{1}{\rho}\frac{\p p}{\p r}= g \frac{dh}{dr}
                \eea
                leading to
                \bea
                UU^{\prime}f^2= -g h^{\prime}+ \frac{\nu U}{h^2}\frac{\p^2 f}{\p 
                \eta^2}
                \eea
                At this point, we need to make a statement about the vertical profile 
                $f(\eta)$.A common situation (constant pressure gradient)is one where 
                the profile is parabolic.This implies (with the boundary conditions 
                f(0)=0,f(1)=1 and $f^(\prime)(1)=0$)that $f(\eta)=2(\eta)-(\eta)^2$.this 
                yields $f^(\prime)^(\prime)=-2$ and we get on the surface (ie.$\eta =1$)
                \bea
                UU^{\prime} = -g h^{\prime}- \frac{2 \nu U}{h^2}
                \eea
                which is the main equation for considering the phenomenon of hydraulic 
                jump.Bohr et.al. arrive at a similar equation.using the continuity 
                conditions as expressed by Eqn(5),
                $U(r) h(r) = \frac{3Q}{4\pi r}$,
                we can write,
                $U^{\prime}(r) h(r)+ U(r) h^{\prime}(r)= - \frac{3Q}{4\pi r^2}$ and 
                this helps us to write Eqn(13)
                \ber
                - \frac{3Q}{4\pi r^2}\frac{U(r)}{h(r)}- \frac{h^{\prime}U^2}{h}&=& -g 
                h^{\prime}- \frac{2 \nu U}{h^2}\nonumber \\
                \frac{dh}{dr}(g- \frac{U^2}{h})&=&[\frac{3Q}{4\pi r^2} - 
                \frac{2\nu}{h}]\frac{U}{h}\nonumber \\
                h \frac{dh}{dr}&=&U\frac{[\frac{3Q}{4\pi r^2} - \frac{2\nu}{h}]}{[g- 
                \frac{U^2}{h}]}
                \eer
                the central equation of our paper.We can write this as a set of two 
                equqtions 
                \ber
                \frac{1}{2}\frac{d h^2}{d \tau}&=&(\frac{3Q}{4\pi r^2}- 
                \frac{2\nu}{h})\frac{3Q}{4\pi rh}\nonumber \\
                \frac{dr}{d\tau}&=& g- \frac{U^2}{h}= g - (\frac{3Q}{4\pi 
                r})^2\frac{1}{h^3}
                \eer
                This system has a fixed point at r=R,$U=U_0$,$h=h_0$ with
                \ber
                U_0^2&=&gh_0(R) \nonumber \\
                h_0(R)&=& \frac{8\pi}{3}\frac{\nu R^2}{Q}
                \eer
                Writing 
                $h=\frac{Q}{\nu}H$,$r=\frac{Q}{\nu}\frac{R}{\sqrt(\frac{8\pi}{3})}$,$u=\sqrt(\frac{Qg}{\nu})U$.
                $U^2=H$ and $H=R^2$.Thus R at the fixed point is:
                $R=\frac{1}{2\pi}(\frac{3}{4})^{\frac{1}{2}}(\frac{8\pi}{3})^{\frac{1}{2}}q^{- 
                \frac{3}{8}}(\nu)^{\frac{5}{8}}g^{-\frac{1}{8}}$. THus in unscaled 
                coordinates,
                $r_(at fixed 
                point)=(0.99)q^{\frac{5}{8}}(\nu)^{-\frac{3}{8}}g^{-\frac{1}{8}}$.It is straightforward to check that the fixed point is astable 
                spiral.
                \par We now return to Eqn(14) and examine the limits $r\gg R$ and $r\ll 
                R$.For $r\gg R$,Eqn(14) simplifies to 
                \bea
                \frac{dh}{dr}\simeq - \frac{3Q\nu}{2\pi g r h^3}
                \eea
                which leads to 
                \bea
                h=[\frac{6}{\pi}\frac{Q\nu}{g}\ln\frac{r_0}{r}]^{1/4}
                \eea
                for $r\gg R$ but $r\ll r_0$.the outer cutoff is $r_0$ beyond which this 
                formula is not valid.We see that h(r) has a very weak r-dependence in 
                this range.for $r\ll R$,on the other hand,we have
                \ber
                \frac{dh}{dr}&\sim& -\frac{\frac{3Q}{4 \pi r h}\frac{3Q}{4\pi 
                r^2}}{(\frac{3Q}{4\pi})^2\frac{1}{h^3 r^2}}\nonumber \\
                &=&-\frac{h}{r}
                \eer
                which integrates to 
                \bea
                rh = constant
                \eea
                In the inner region the height drops off as $h(r)\sim \frac{1}{r}$ and 
                in the outer region it is approximartely constant.The dividing point of 
                the flow is at r=R,which is where the jump takes place so that falling 
                $h\sim  \frac{1}{r}$ profile can meet the nearly constant $h\sim (ln 
                \frac{1}{r})^{\frac{1}{4}}$ profile.
                \par This leads to the jump radius being 
                \bea
                R = (\frac{243}{256})^{1/8}\frac{Q}{2\pi}^{5/8}\nu^{-3/8}g^{-1/8}
                \eea
                This relation is very similar to that found by Bohr et.al. The 
                difference in the prefactor comes from the difference between the choice of the 
                z-dependence of the radial velocity.Bohr et.al.
                chose to work with a z-averaged velocity while we have demonstrated the 
                jump by studying the radial flow velocity at the free surface.

                \end{document}